# Versatile and reconfigurable integrated silicon nitride photonic microresonator

Tong Lin,[a,b,†,*] Haoran Wang,[a,†] Xiaoyu Hao,[a,†] Liu Li,[c] Ziyang Xiong,[a] Hao Deng,[a] Yan Fan,[a] Shihua Chen,[d] Junpeng Lu[a], Zhenhua Ni[a,d]

[a]Southeast University, Photonic integrated circuits & systems group, School of Electronic Science and Engineering, No. 2 Southeast University Road, Jiangning District, Nanjing, China

[b]Key Lab of Modern Optical Technologies of Education Ministry of China, Soochow University, No.1 Shizi Street, Suzhou, China

[c]Southeast University, School of Material Science and Engineering, No. 2 Southeast University Road, Jiangning District, Nanjing, China

[d]Southeast University, Advanced Ocean Institute, No. 2 Southeast University Road, Jiangning District, Nantong, China

**Abstract**. Unlocking the full potential of integrated photonics requires versatile, multi-functional devices that can adapt to diverse application demands. However, confronting this challenge with conventional single-function resonators often results in tedious and complex systems. We present an elegant solution: a versatile and reconfigurable dual-polarization $Si_3N_4$ microresonator that represents a paradigm shift in on-chip photonic designs. Our device, based on a binary-star orbital architecture, can be dynamically reconfigured into three distinct topologies: a Möbius-like microcavity, a Fabry-Pérot resonator, and a microring resonator. This unprecedented functionality is enabled by a tunable balanced Mach-Zehnder interferometer that facilitates controllable mutual mode coupling of counterpropagating lights using a single control knob. We experimentally demonstrate that the device not only supports polarization-diverse operation on a compact footprint but also gives rise to a rich variety of physical phenomena, including a standing wave cavity, a traveling wave cavity, free spectral range multiplication, and the photonic pinning effect. These behaviors are accurately modeled using the Transfer Matrix Method and intuitively explained by Temporal Coupled Mode Theory. Our results underscore the profound potential for a chip-scale platform to realize reconfigurable reconstructive spectrometers and on-chip synthetic dimensions for topological physics.

[†]Tong Lin, Haoran Wang, and Xiaoyu Hao contributed equally to this work.
[*]Tong Lin, E-mail:lintong@seu.edu.cn

# 1 Introduction

Integrated photonic microresonators[1] have become a cornerstone technology for the miniaturization and mass production of optical components, with broad applications in high-speed optical communications,[2,3] photonic computing,[4,5] chemical sensing,[6] biomedical diagnostics,[7] and quantum information processing.[8] Their appeal lies in compact form factors, high quality factors, and strong light–matter interactions. However, the field remains constrained by a fundamental limitation: conventional photonic resonators are inherently single-function devices. Without extensive circuit-level redesign, they cannot adapt to varying application demands. This architectural rigidity hinders the development of versatile on-chip systems that require reconfigurable spectral control, multifunctional operation, and polarization diversity within a minimal footprint.

In recent years, significant research efforts have focused on developing reconfigurable integrated microresonators, primarily leveraging interferometric designs to manipulate optical pathways. In bulk optical systems, aligning even a few interferometric elements remains challenging. Advances in micro/nanofabrication now enable the monolithic integration of complex photonic circuits, leading to diverse approaches such as: reconfigurable optical gates based on hexagonal cells,[1,9] double-injection resonators,[10,11] networks of microring resonators (MRRs),[12,13] Mach–Zehnder interferometer (MZI)-assisted ring resonators,[14,15] and interleaver arrays.[16] Despite enabling sophisticated spectral shaping, these methods universally rely on intricate circuit combinations and multi-channel control schemes—resulting in cumbersome designs and computationally intensive operation. A fundamentally distinct solution—a versatile microresonator with single-knob reconfigurability—is thus critically needed to advance programmable photonics.

In this work, we bridge this critical gap by introducing a versatile reconfigurable dual-polarization silicon nitride ($Si_3N_4$) microresonator based on a binary-star orbital architecture. We demonstrate that this single, compact device can be dynamically reconfigured into three distinct operational modes—a Möbius-like microcavity, a Fabry-Pérot resonator, and a MRR—enabling unprecedented multi-functionality on chip using a single control knob. Furthermore, our design benefits from the use of an 0.8-μm thick $Si_3N_4$ photonic platform, which is engineered to minimize birefringence and support polarization diversity. The cavity dynamics are accurately modeled by the transfer matrix method (TMM) and intuitively explained by the temporal coupled mode theory (TCMT). Through experiment validations, we further unveil a suite of intriguing physical phenomena intrinsic to this platform, including free spectral range (FSR) multiplication and the photonic pinning effect. Our findings not only provide a versatile building block for programmable photonic circuits but also open new pathways toward on-chip synthetic dimensions and reconfigurable spectroscopic systems.[17]

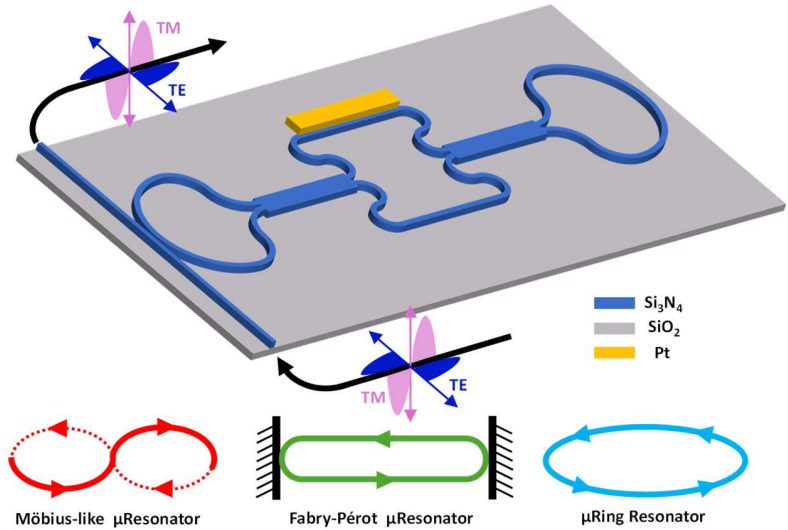

**Fig. 1** The artistic view of the proposed reconfigurable dual-polarization $Si_3N_4$ microresonator, supporting three distinctive resonators.

## 2  Reconfigurable Photonic Microresonator

We present a reconfigurable dual-polarization $Si_3N_4$ microresonator, leveraging a novel optical architecture capable of dynamically switching between three distinct resonator configurations. As illustrated in Fig. 1, the proposed device features a resonant loop resembling a binary-star orbital shape, side-coupled to a bus waveguide for optical input and output. This loop is formed by interconnecting two identical Sagnac-loop retroreflectors in a port-to-port configuration. Each retroreflector is constructed using a 2×2 multimode interferometer (MMI) whose two output ports are looped back to form a closed path. Phase control is achieved via the thermo-optic effect using an integrated microheater atop one of the connecting waveguides. The core of the microresonator can also be viewed as a tunable balanced MZI, which facilitates controllable mutual mode coupling. The resulting optical paths—incorporating intricate reflection and transmission dynamics within the resonant structure—give rise to a rich variety of physical phenomena, i.e., standing wave cavity, traveling wave cavity, FSR multiplication, and photonic pinning effect. Fabricated on a 0.8-μm-thick Si$_3$N$_4$-on-insulator platform,[18] the Si$_3$N$_4$ microresonator enables a straightforward polarization-diverse operation in a compact manner, making it suitable for advanced integrated photonic applications.

*2.1 Architecture of the Photonic Router*

To achieve dual-polarization reconfigurability, we implement a tunable Sagnac retroreflector as the core on-chip photonic router, leveraging polarization insensitive interferometric design. Figure 2(a) schematically illustrates this router, which employs polarization insensitive MMIs and a thermo-optic phase shifter to dynamically regulate optical paths. Structurally, the photonic router functions as a balanced MZI with interconnected outputs—akin to a drop shape with a perimeter of about 634.16 μm ($L_1$). Within the MZI, two 2×2 MMIs are interconnected port-to-port via S-bends, each spanning 634.92 μm ($L_2$) with four 180° circular bends (25 μm radius). The MMI operates on the self-imaging principle to equally split incident light. We optimize MMI performance using finite-difference time-domain (FDTD) simulations (Ansys Lumerical Inc.). Equalizing the beat lengths for both polarizations yields polarization-insensitive operation with a central multimode body cross-section of 130 μm×12.6 μm. At multimode interfaces, pairs of 3-μm-wide waveguides (1.2 μm lateral spacing) are adiabatically tapered to 1 μm width over 40 μm lengths longitudinally. While non-tapered waveguides maintain 1 μm width, the entire waveguide layer is 0.8 μm thick.

Figures 2(b)–(c) validate near-identical MMI output transmissions for TE and TM modes. At 1550 nm, both output branches achieve nearly 49% transmission efficiency for TE polarization, while TM polarization exhibits slightly higher insertion loss with peak efficiency at around 47.5%. Using Ansys Lumerical INTERCONNECT, we simulate the Sagnac retroreflector's tunable response. Figures 2(d–e) plot normalized transmission (solid blue) and reflection (dashed blue) spectra versus phase shift $\Delta\phi$. For TE input (Fig. 2(d)), the retroreflector operation occurs at $\Delta\phi=0.5\pi$, where reflection approaches unity and transmission minimizes. For TM input (Fig. 2(e)), peak reflection reduces to 80% at $\Delta\phi=0.5\pi$ due to MMI insertion loss. At $\Delta\phi=0$ or $\pi$, the device functions as an all-pass filter with near 100% transmission through the lower port. As the wide range phase shift is readily attained with rigorous thermal-optic tuning, these simulation results confirm tunable routing capability of the constructed retroreflector.

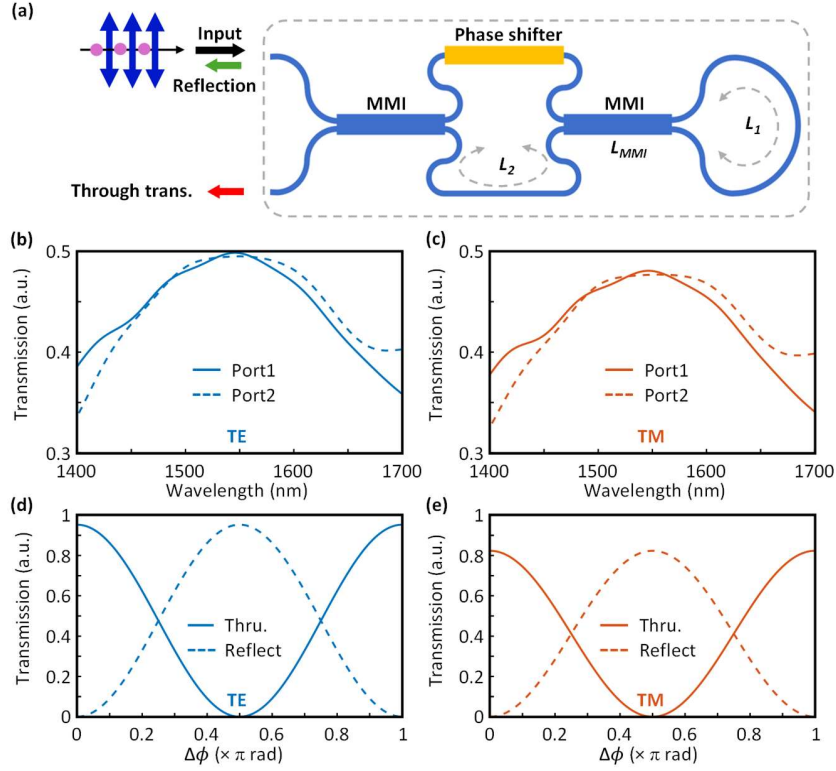

**Fig. 2** (a) The schematic of the tunable Sagnac retro-reflector as the photonic router ($L_{MMI}$=130 μm). The two port transmission spectra of a polarization-insensitive MMI for (b) TE mode; (c) TM mode. The reflection and transmission spectra of the constructed Sagnac retro-reflector for (d) TE mode; (e) TM mode.

## 3 Experimental Results and Discussion

*3.1 Experimental Implementation*

We experimentally demonstrate the reconfigurability of the fabricated photonic device for both TE and TM polarizations. The device was manufactured on a commercial $Si_3N_4$ photonics platform (Ligentec Inc.). A microscope image of the structure is provided in the top inset of Fig. 3. The $Si_3N_4$ waveguide features nearly vertical sidewalls (89° sidewall angle), owing to a well-controlled etching process. With a cross-section of 1 μm×0.8 μm (except in the MMI multimode regions), the $Si_3N_4$ waveguide exhibits negligible birefringence. For optical characterization, light from a tunable laser is coupled into the device under test (DUT) via a lensed fiber, as depicted in Fig. 3. The output light is collimated using an aspheric lens and passed through a free-space polarizer. By rotating the polarizer, we selectively measure the transmission spectra for TE and TM polarizations using a photodetector. Reconfigurability is achieved via integrated microheaters, positioned 3.3 µm above the waveguides, enabling efficient phase tuning of the photonic router.

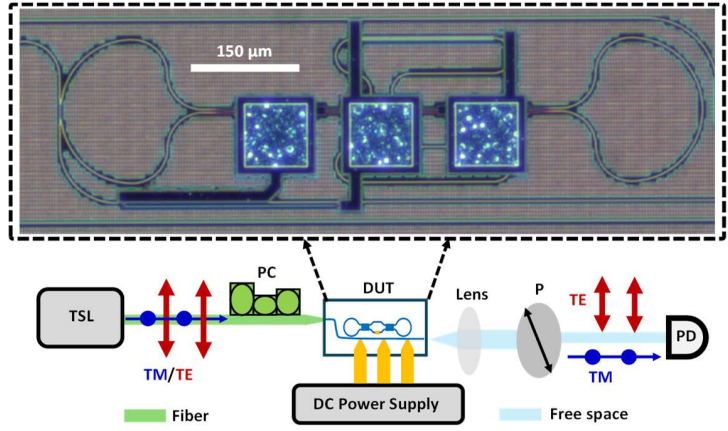

**Fig. 3** The optical microscope image of the fabricated device and the experimental apparatus for characterizing its optical properties: TSL (Santec TLS-570), PC (Polarization controller), DC power supply (Keysight 33522B), P (OPPF05-NIR, JCOPTIX), PD (Santec MPM-210).

*3.2 Reconfigurability and Spectral Evolution*

We experimentally demonstrate the reconfigurable functionality of the device for both TE and TM polarizations. The observed behavior is well captured by TMM simulations (see Sec. S1 in the Supplementary Material for details). The reconfiguration mechanism relies critically on the tunable reflection introduced by the photonic router. From experimental transmission spectra, we extract the normalized reflectivity as a function of the applied phase shift $\Delta\phi$, which exhibits a period of $\pi$. For TE-polarized input, nine distinct states are identified, corresponding to the labeled points on the reflection curve in the top-right panel of Fig. 4. In State I, the resonator operates in a Möbius-like topology due to the cross-intersection behavior of the photonic router at zero phase shift. This unidirectional loop structure, with its characteristic one-sided topology, defines an optical path length $L$ ($L=2L_1+2L_2+4L_{MMI}$), yielding a FSR of about 0.3389 nm, consistent with the theoretical value: $FSR=c/(n_gL)$. The presence of non-zero bias power results from fabrication-induced arm imbalance in the interferometer, a common feature in integrated MZIs.[19] As the applied power increases from State II to State IV, enhanced reflection from the photonic router leads to progressively larger mode splitting. At peak reflectivity, the router acts as a bidirectional retroreflector, effectively emulating a standing-wave Fabry–Pérot cavity with two end mirrors— as illustrated in the green inset. The corresponding FSR is halved to 0.1697 nm, matching the canonical Fabry–Pérot formula: *FSR=c/(2n_gL)*. Beyond 153.7 mW, reflectivity decreases monotonically, reducing the mode splitting through States VI to VIII. Notably, at a π phase shift, the photonic router enters the "bar" state, enabling microring-type resonance. In State IX, the measured FSR is approximately 0.3345 nm—nearly identical to the Möbius-like case—as the optical path length remains unchanged. Variations in extinction ratios (ERs) across states are attributed to insertion loss differences between the "on" and "off" states of the MZI, arising primarily from the 2×2 MMI imbalance.

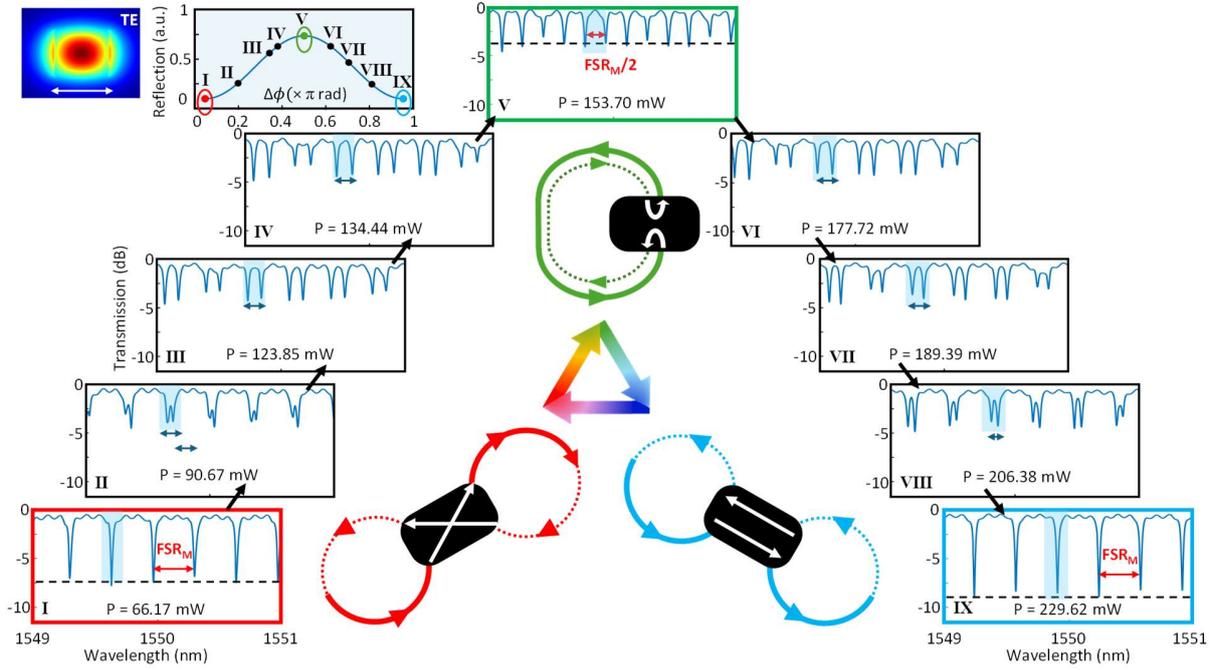

**Fig. 4** The spectral evolutions of the reconfigurable $Si_3N_4$ microresonator for TE-polarized input with the wavelength spanning from 1549 nm to 1551 nm by altering the applied electrical power (*P*). The top left panel shows the simulated fundamental TE mode profile ($n_g$=2.12124) and the normalized reflectivity versus the applied phase shift, derived from all the measures spectra.

Our measurements for TM-polarized input confirm the same full range of reconfigurable functionality, providing compelling evidence of the device's robust performance and polarization-insensitive operation. Similarly, Figure 5 depicts the transmission spectral evolutions as the electrical tuning power increases from 44.5 mW to 216.45 mW. In State I, the resonator exhibits Möbius-like behavior with a FSR of approximately 0.3352 nm. The slight discrepancy between TE and TM FSR values stems from a group index difference of $6\times10^{-3}$ between the two fundamental modes. In State V, the Fabry–Pérot resonance becomes dominant, exhibiting a nearly halved FSR of about 0.1706 nm. The thermal tuning power slightly deviates from the value required for peak reflection, resulting in a minor FSR offset of only 5 pm. The dynamic evolution stops manually in State IX, where the resonator is reconfigured as a MRR with a well-defined FSR of about 0.3337 nm and comparatively higher ERs, consistent with the behavior explained earlier. Interestingly, the loaded Q-factors for the TM mode are higher than those for TE, despite the larger simulated insertion loss of the TM-polarized MMI (Figs. 2(b–c)). We attribute this enhancement to the lower propagation loss of the TM mode.

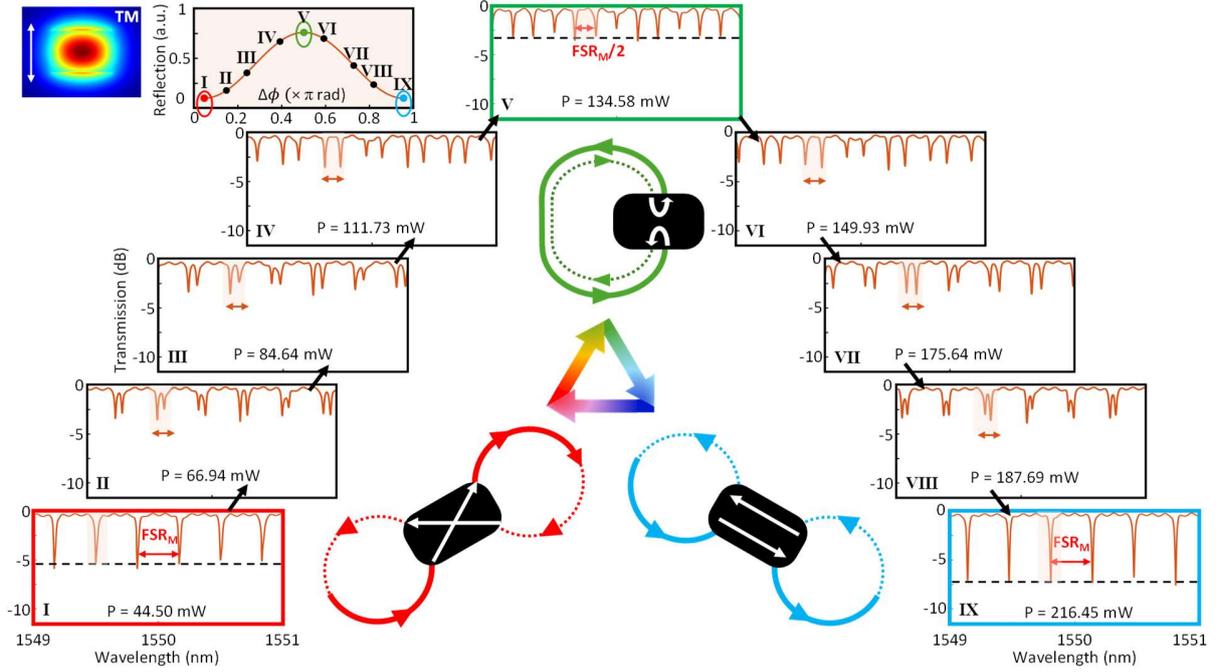

**Fig. 5** The spectral evolutions of the reconfigurable $Si_3N_4$ microresonator for the TM-polarized input with the wavelength spanning from 1549 nm to 1551 nm by altering the applied electrical power ($P$). The top left panel shows the simulated fundamental TM mode profile ($n_g$=2.12729) and the normalized reflectivity versus the applied phase shift, derived from all the measures spectra.

*3.3 Mode Splitting and Avoided Mode Crossing*

To gain a more insightful understanding of the underlying physics governing the spectral evolution and mode splitting, we perform comprehensive analysis using the TCMT. As a more fundamental view of the light interaction, this model focuses on the coupling between clockwise (CW) and counterclockwise (CCW) traveling waves within the chip-scale resonator. For any non-zero reflectivity (where $\Delta\phi$ is not an integer multiple of $\pi$), CW and CCW waves coexist within the resonator. These two modes travel the same path and can be converted into each other via the phase difference between the two arms of the balanced MZI, as illustrated in Fig. 6(a). Due to the structural symmetry, the impact of the CW and CCW waves on the transmission spectrum is equivalent to the coupling of dual rings of equal length. The coupling coefficient (*k*) is equivalent to the normalized reflectivity *R* of the Sagnac retroreflector (i.e., $K=R(\lambda)/R_{max}$). When *k*=0, the two MRRs are completely decoupled, corresponding to the case where $\Delta\phi$ is an integer multiple of π. When *k*=1, the two MRRs are fully coupled, and the optical path length becomes twice the original, resulting in a halved FSR when $\Delta\phi=\pi/2$. The mode splitting in intermediate states can be quantitatively analyzed using the two-ring mode coupling theory:

$$\omega_n^{1,2} = \omega_n - i\gamma \pm \frac{\theta_n}{T_{RT}}, \tag{1}$$

where $\omega_n^{1,2}$ represent the two super modes corresponding to the $n_{th}$ azimuthal mode number, $\omega_n$ is the center angular frequency of this longitudinal mode, $\theta_n$ is the coupling rotation angle, related to the power coupling coefficient $k_n$ by $\theta_n=\arcsin(k_n)$, and $T_{RT}$ denotes the cavity round-trip time. More details are be found in S2 in the Supplementary Material.

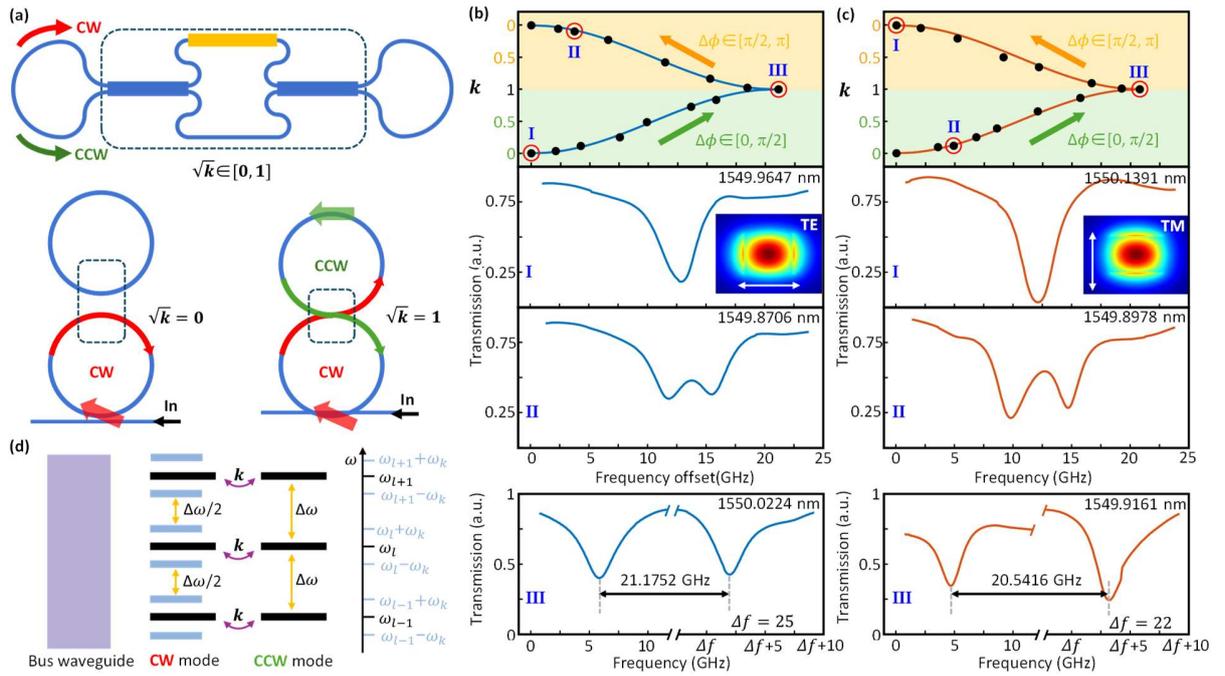

**Fig. 6** (a) The schematic of the tunable dual-cavity system supporting CW and CCW modes; the equivalent dual-ring system when *k*=0 or 1. (b) Measured fundamantal TE resonant frequency difference (black points) with theoretical curve fit (blue line). Distinctive resonance profiles corresponding to States I, II, and III. (c) Measured fundamantal TM resonant frequency difference (black points) with theoretical curve fit (orange line). Distinctive resonance profiles corresponding to States I, II, and III. (d) Energy-level diagram illustration of the mode splitting with the largest coupling.

We validate that the experimental mode splitting evolution for both polarizations during device reconfiguration aligns precisely with the TCMT predictions. As shown in Fig. 6(b), the resonant frequency difference exhibits a direct dependence on the inter-ring coupling strength ($k$). At zero coupling (State I), a single resonance dip occurs near 1549.96 nm with the typical Lorentzian lineshape. With increasing $k$ in State II, hybridized supermodes emerge through enhanced CW-CCW mode coupling, generating observable mode splitting. This culminates in State III ($\Delta\phi = \pi/2$, $k$ is maximized), where maximal splitting yields two distinct resonances separated by approximately 21.175 GHz – effectively halving the FSR in State I. Similar behavior is observed for TM polarization in Fig. 6(c), with a maximum frequency splitting of 20.541 GHz. This tunable dual-mode system constitutes a prevailing photonic molecule,[20] demonstrating analogs to canonical two-level quantum systems such as Autler-Townes splitting. Figure 6(d) accordingly illustrates the corresponding energy-level diagram for CW and CCW modes,[21] with the dynamically controlled spectral response establishing a reconfigurable platform that emulates artificial atom- and molecule-like systems.

We demonstrate the dual polarization photonic pinning effect enabled by phase-compensated tunable coupling. In conventional mechanical tuning methods,[22, 23] which modulate coupling strength solely through inter-resonator gap adjustment, the hybridized even and odd supermodes shift simultaneously in opposite directions within the frequency domain.[24] By contrast, our interferometric coupler tuning approach introduces a compensatory phase term that precisely counteracts mode splitting dynamics. Consequently, one supermode remains frequency-pinned, as validated theoretically and experimentally below (See S2 in the Supplementary Material for more details). Figures 7(a–b) present numerically simulated relative resonant wavelengths of the two supermodes as a function of the applied phase difference $\Delta\phi$ (in units of $\pi$ radians) under TE and TM polarized inputs. For $0<\Delta\phi<0.5\pi$, the odd supermode (out-of-phase state) remains pinned at a fixed wavelength, while the resonant wavelength of the even supermode (in-phase state) shifts linearly with $\Delta\phi$. For $0.5\pi<\Delta\phi<\pi$, the behavior of the two supermodes is reversed: the odd supermode now shifts linearly with the phase, while the even supermode becomes wavelength-pinned.

In addition to phase-sensitive coupling variations, the thermal tuning approach introduces thermal crosstalk,[25] that inevitably increasing the cavity length and causing a red shift in the spectra. The measured resonant wavelength offsets under varying applied electrical power for TE and TM polarizations are shown in Figs. 7(c-d). The actual trendlines of photonic pinned supermodes are slightly titled. Yellow lines represent the linear least-squares fits to the experimental data (red solid dots). Therefore, the thermal crosstalk values are derived as 1.718 pm/mW for TE polarization and 1.747 pm/mW for TM polarization, indicating consistent and polarization-insensitive thermal response. This pinning phenomenon is analogous to avoided mode crossing (AMX)-induced Kerr comb generation.[26, 27] By controlling AMX location, initial comb lines can be selectively generated at specified resonances, facilitating repetition-rate-selectable combs. As the representative behavior of photonic molecules,[3, 28] the AMX has also been widely explored in both classical and quantum scenarios.[29-31] The observed photonic pinning effect underscores the rich physical dynamics underlying our reconfigurable microresonator.

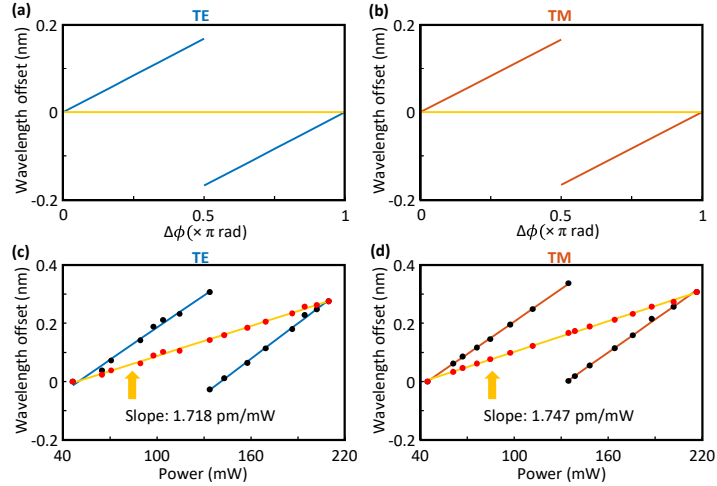

**Fig. 7** Wavelength offset of the resonant supermodes as a function of the applied phase difference (Δφ, in units of π radians) for (a) TE and (b) TM polarized inputs. Experimental resonance wavelength shift under varying applied power for (c) TE and (d) TM polarizations. Data points (black and red circles) correspond to measured resonance (even and odd modes) positions, and solid lines (blue/orange and yellow) indicate linear fits.

## 4    Conclusion

To sum up, we propose and experimentally demonstrate a versatile reconfigurable integrated $Si_3N_4$ photonic microresonator that fundamentally expands the capabilities of on-chip devices. We successfully show that a single, compact device can dynamically transition through three major operational modes —a Möbius-like cavity, a Fabry-Pérot resonator, and a standard microring resonator—a behavior that is accurately modeled using the TMM and intuitively explained by the TCMT. Beyond comprehensive demonstration of reconfigurability, our experiments reveal several surprising phenomena. We show that the device can achieve FSR multiplication by leveraging a unique photonic router topology, where the FSR is halved as a direct result of twice the optical path before constructive interference. Furthermore, we observe a dual-polarization photonic pinning effect enabled by phase-compensated tuning, which allows one supermode to remain frequency-pinned. This phenomenon has profound implications for photonic systems, demonstrating analogs to canonical two-level quantum systems and offering a route for realizing artificial atom- and molecule-like systems[32] and synthetic dimensions[33-37] on a chip. The demonstrated ability to achieve multi-functionality and rich physical phenomena on a single platform represents a significant step towards advanced programmable photonic system-on-chip.

**Disclosures**

The authors declare no conflicts of interest.

**Code and Data Availability**

The data that support the findings of this study are available from the corresponding author on reasonable request.

*Acknowledgments*

This research was supported by the National Natural Science Foundation of China (NSFC) (Grant Nos. 62105061, 12374301, and 62225404); Jiangsu Provincial Frontier Technology Research and Development Program (Grant No. BF2024070); the National Key R&D Program of China (Grant No. 2024YFA1210500); the Key Lab of Modern Optical Technologies of Education, Ministry of China, Soochow University. The authors thank Yong Xu for insightful discussion.

**Tong Lin** is an associate professor at the Southeast University. He received his BS degree in optics from the Zhejiang University in 2012, and his PhD degree in mechanical engineering from the National University of Singapore in 2016. He is the author of more than 50 journal papers. His current research interests include silicon nitride photonics, frequency comb, optical interconnects, silicon photonics, metrology, and optoelectronic systems. He is a member of SPIE.

Biographies and photographs for the other authors are not available.

## Caption List

**Fig. 1** The artistic view of the proposed reconfigurable dual-polarization Si3N4 microresonator, supporting three distinctive resonators.

**Fig. 2** (a) The schematic of the tunable Sagnac retro-reflector as the photonic router ($L_{MMI}$=130 μm). The two port transmission spectra of a polarization-insensitive MMI for (b) TE mode; (c) TM mode. The reflection and transmission spectra of the constructed Sagnac retro-reflector for (d) TE mode; (e) TM mode.

**Fig. 3** The optical microscope image of the fabricated device and the experimental apparatus for characterizing its optical properties: TSL (Santec TLS-570), PC (Polarization controller), DC power supply (Keysight 33522B), P (OPPF05-NIR, JCOPTIX), PD (Santec MPM-210).

**Fig. 4** The spectral evolutions of the reconfigurable $Si_3N_4$ microresonator for TE-polarized input with the wavelength spanning from 1549 nm to 1551 nm by altering the applied electrical power (P). The top left panel shows the simulated fundamental TE mode profile ($n_g$=2.12124) and the normalized reflectivity versus the applied phase shift, derived from all the measures spectra.

**Fig. 5** The spectral evolutions of the reconfigurable $Si_3N_4$ microresonator for the TM-polarized input with the wavelength spanning from 1549 nm to 1551 nm by altering the applied electrical power (P). The top left panel shows the simulated fundamental TM mode profile ($n_g$=2.12729) and the normalized reflectivity versus the applied phase shift, derived from all the measures spectra.

**Fig. 6** (a) The schematic of the tunable dual-cavity system supporting CW and CCW modes; the equivalent dual-ring system when $k$=0 or 1. (b) Measured fundamantal TE resonant frequency difference (black points) with theoretical curve fit (blue line). Distinctive resonance profiles corresponding to States I, II, and III. (c) Measured fundamantal TM resonant frequency difference (black points) with theoretical curve fit (orange line). Distinctive resonance profiles corresponding to States I, II, and III. (d) Energy-level diagram illustration of the mode splitting with the largest coupling.

**Fig. 7** Wavelength offset of the resonant supermodes as a function of the applied phase difference (Δφ, in units of π radians) for (a) TE and (b) TM polarized inputs. Experimental resonance wavelength shift under varying applied power for (c) TE and (d) TM polarizations. Data points

(black and red circles) correspond to measured resonance (even and odd modes) positions, and solid lines (blue/orange and yellow) indicate linear fits.